\documentclass[format=sigconf,natbib=true, anonymous=false]{acmart}

\usepackage{nicefrac}
\usepackage{multirow}
\usepackage{xspace}
\usepackage{booktabs}
\usepackage{makecell}
\usepackage{moresize}
\usepackage{tabularx}
\usepackage{adjustbox}
\usepackage[normalem]{ulem}
\usepackage{siunitx}
\usepackage{tikz}
\usepackage{amsthm}
\usepackage{graphicx}
\usepackage[show]{chato-notes}
\usepackage{soul}

\AtBeginDocument{%
  \providecommand\BibTeX{{%
    \normalfont B\kern-0.5em{\scshape i\kern-0.25em b}\kern-0.8em\TeX}}}

\copyrightyear{2022} 
\acmYear{2022} 
\setcopyright{acmlicensed}
\acmConference[SIGIR '22]{Proceedings of the 45th International ACM SIGIR Conference on Research and Development in Information Retrieval}{July 11--15, 2022}{Madrid, Spain}
\acmBooktitle{Proceedings of the 45th International ACM SIGIR Conference on Research and Development in Information Retrieval (SIGIR '22), July 11--15, 2022, Madrid, Spain}
\acmPrice{15.00}
\acmDOI{10.1145/3477495.3531774}
\acmISBN{978-1-4503-8732-3/22/07}

\newcommand{\maxscore}{\textsf{MaxScore}}
\newcommand{\wand}{\textsf{WAND}}
\newcommand{\bmw}{\textsf{BMW}}

\newcommand{\bmtf}{\textsf{BM25}\xspace}
\newcommand{\doctfquery}{\textsf{DocT5Query}\xspace}
\newcommand{\deepct}{\textsf{DeepCT}\xspace}

\newcommand{\deepimpact}{\textsf{DeepImpact}}

\newcommand{\unicoil}{\textsf{uniCOIL}}
\newcommand{\tildetwo}{\textsf{TILDEv2}}
\newcommand{\spladetwo}{\textsf{SPLADEv2}}
\newcommand{\gt}{\textsf{DeepImpact-GT}}
\newcommand{\gti}{\textsf{DeepImpact-GTI}}

\newcommand{\msmarcodev}{MSMARCO Dev Queries\xspace}
\newcommand{\trecdl}{\textsf{TREC 2019}\xspace}
\newcommand{\trecdltw}{\textsf{TREC 2020}\xspace}

\newcommand{\multibert}{\textsf{DeepImpact}}

\newcommand{\myparagraph}[1]{\paragraph*{\hspace*{-\parindent}\normalsize\bf#1}}
\newcommand{\mycaption}[1]{\caption{{\rm{#1}}}}

\begin{document}
 \fancyhead{}
\title{Faster Learned Sparse Retrieval with Guided Traversal}

\settopmatter{printacmref=true}
\author{Antonio Mallia}
\email{antonio.mallia@nyu.edu}
\affiliation{
  \institution{New York University}
  \city{New York}
  \country{US}
}

\author{Joel Mackenzie}
\email{joel.mackenzie@uq.edu.au}
\affiliation{
 \institution{The University of Queensland}
 \city{Brisbane}
 \country{Australia}
}
\author{Torsten Suel}
\email{torsten.suel@nyu.edu}
\affiliation{
  \institution{New York University}
    \city{New York}
  \country{US}
}
\author{Nicola Tonellotto}
\email{nicola.tonellotto@unipi.it}
\affiliation{
  \institution{University of Pisa}
    \city{Pisa}
  \country{Italy}
}

\renewcommand{\shortauthors}{Trovato and Tobin, et al.}

\begin{abstract}
Neural information retrieval architectures based on transformers such as BERT are able to significantly improve system effectiveness over traditional sparse models such as {BM25}.
Though highly effective, these neural approaches are very expensive to run, making them difficult to deploy under strict latency constraints.
To address this limitation, recent studies have proposed new families of learned sparse models that try to match the effectiveness of learned dense models, %
while leveraging the traditional inverted index data structure for efficiency.

Current learned sparse models learn the weights of terms in documents and, sometimes, queries; however, they exploit different vocabulary structures, document expansion techniques, and query expansion strategies, which can make them slower than traditional sparse models such as
{BM25}.
In this work, we propose a novel indexing and query processing technique that exploits a traditional sparse model's ``guidance'' to efficiently traverse the index, allowing the more effective learned model to execute fewer
scoring operations. %
Our experiments show that our guided processing heuristic is able to boost the efficiency of the underlying learned sparse model by a factor of four without any measurable loss of effectiveness.
 
\end{abstract}

\ccsdesc[500]{Information systems~Information retrieval query processing}
\ccsdesc[500]{Information systems~Retrieval models and ranking}
\ccsdesc[500]{Information systems~Search engine architectures and scalability}
 \keywords{Query processing, Inverted Index, Learned Sparse Retrieval}

\maketitle

\section{Introduction}\label{sec-intro}

Neural information retrieval architectures based on transformers such as BERT~\cite{devlin2018bert} are able to significantly improve system effectiveness over traditional sparse models such as {BM25}. At the same time, they also pose new challenges, as transformers are very computationally expensive. This has motivated a lot of recent work on making transformer-based ranking more efficient. This includes the use of special-purpose hardware such as Google's Tensor Processing Units (TPUs) \cite{tpu}, nearest-neighbor methods for quickly identifying candidate documents in dense retrieval scenarios \cite{JDH17}, and the design of late-interaction transformer-based rankers such as ColBERT \cite{colbert}.

A straightforward application of transformer-based ranking involves applying a transformer at query time to each document that is being re-ranked, leading to significant computational costs. While this can be reduced through approaches such as ColBERT~\cite{colbert}, or by re-ranking only a small set of candidates, the resulting methods are still much more expensive than a simple ranking function that can be directly evaluated over an inverted index. 
On the other hand, approaches based on learned sparse representations aim to come close to the best transformer-based methods in effectiveness while preserving the efficiency of simple bag-of-words rankers such as {BM25}~\cite{bm25}.

The main goal in (most) learned sparse representations, as discussed by \citet{unicoil}, is to learn a set of terms under which a document should be indexed (document expansion), and the impact scores that should be stored in the corresponding inverted index postings (learning impacts), such that the resulting ranking function approximates the effectiveness of a full transformer-based ranker while retaining the efficiency of the fastest inverted-index based methods. Recent work has shown, however, that while effective retrieval is possible with learned sparse approaches, they are often still much slower than their traditional counterparts {\cite{deepct-efficiency, deepimpactv1, wacky}}. In this work, we propose a novel heuristic index traversal mechanism that closes the performance gap between learned and traditional rankers. Experiments over the MSMARCO passage collection demonstrate that our heuristic approach can accelerate {\multibert} retrieval by a factor of four without any measurable loss in effectiveness.

\section{Background \& Motivation}\label{sec:background}
We now briefly introduce some background and related work before motivating our proposed approach.

\myparagraph{Learned Sparse Models}
While various learned sparse models have been proposed, they all improve effectiveness in two ways:
\begin{enumerate}
    \item {\emph{Term Expansion:}} Adding {\emph{new}} terms to document or query representations; and
    \item {\emph{Term Re-weighting:}} Changing the {\emph{weights}} of the terms in documents or queries.
\end{enumerate}
To this end, learned sparse models typically make use of the contextual word representations of large language models such as BERT, allowing them to ``learn'' when new terms should be added, or when terms are important.
It is important to note that both expansion and re-weighting can be applied to documents (before indexing), to queries (before searching), or any combination thereof. 

\deepct~\cite{deepct,deepct2} is the first example of learned sparse retrieval, exploiting the contextual word representations from BERT to re-weight term frequencies for {BM25} scoring. 
The main limitation of \deepct\ lies in the fact that it does not address the vocabulary mismatch problem~\cite{zhao2012modeling}: only terms already appearing in the documents will receive learned weights to improve their relevance signals. 
{\doctfquery} {\cite{docTTTTTquery}} addresses the vocabulary mismatch problem by expanding documents offline via the T5 {\cite{tfive}} sequence-to-sequence model. However, {\doctfquery} does not explicitly re-weight the terms within each document and, like {\deepct}, relies on {BM25} scoring.

A way to address the shortcomings of {\deepct} and {\doctfquery} was first proposed by \citet{deepimpactv1}. Instead of using the original document collection, their model, {\deepimpact}, expands the documents to include new terms to address the vocabulary mismatch via the {\doctfquery} model. Then, instead of scoring the documents with {BM25}, {\multibert} directly optimizes the sum of query term impacts to maximize the score difference between relevant and non-relevant passages for the query. That is, {\deepimpact} actually {\emph{learns}} the ranking function instead of applying an existing one. %

Other approaches for learning expansion terms and term importance re-weighting include \unicoil{}~\cite{unicoil}, {\tildetwo}~\cite{tildetwo}, and \spladetwo{}~\cite{splade}, all of which employ document expansion and the use of a contextualized language model to learn a ranking model. As opposed to \multibert{}, \unicoil\ and \spladetwo\ also perform weighting on the query terms, such that document ranking becomes a {\emph{weighted}} sum over term impacts. Furthermore, \spladetwo\ also employs query expansion, thereby adding new terms to each query rather than just re-weighting the existing ones.

In order to make use of an inverted index for querying, all of the learned term scores derived from the aforementioned models are quantized and embedded into the postings lists of the index. Then querying proceeds as usual, with document scores computed as the weighted sum of term impacts. Although {\tildetwo} was initially proposed in the context of re-ranking a set of candidates, we employ it here as an effective first-stage ranker. It is also worth mentioning that \citet{splade} train \spladetwo\ using knowledge distillation~\cite{hofstatter2020improving}, an approach which substantially increases the model accuracy with no particular effect on the retrieval latency; this could be leveraged by the other models as well, but we leave this as future work.

Table~\ref{tab:collections} summarizes the index statistics and the average query lengths of the different learned sparse models. The average number of postings per term ranges from 100 to 178 for methods exploiting word-level tokenization, namely \bmtf, \deepct, \doctfquery, and \deepimpact, and from 21K to 72K postings per term for methods exploiting BERT's WordPiece tokens~\cite{devlin2018bert}, namely \unicoil, \tildetwo, and \spladetwo. 
The latter tokenization algorithm greedily restricts the number of tokens to subwords that are part of the BERT vocabulary. This process is very different from the more popular word-level tokenization, where tokens are generated by splitting on whitespaces and punctuation characters.     
These ranges, as well as the average number of query terms, have a serious impact on the query processing times, as we will show in Section~\ref{sec:exp}.

\begin{table}[h]
\mycaption{Basic statistics for the indexes of several traditional and learned sparse retrieval models.}
\centering
\begin{tabular}{l r r r r r r}
\toprule
\multicolumn{1}{c}{Model}  & \multicolumn{1}{c}{Terms} & \multicolumn{1}{c}{Postings}  & \multicolumn{1}{c}{Avg. Query Length} \\
\midrule
\bmtf    & \num{2660824}    & \num{266247718} & 4.5\\
\deepct   & \num{989873}    & \num{128969826} & 4.5\\
\doctfquery        & \num{3929111}    & \num{452197951} & 4.5 \\
\unicoil        & \num{27678}    & \num{587435995} & 686.3 \\
\tildetwo        & \num{27437}    & \num{809658361} & 4.9 \\
\spladetwo        & \num{28131}    & \num{2028512653} & 2037.8 \\
\multibert        & \num{3514102}    & \num{628412657} & 4.2  \\

\bottomrule
\end{tabular}
\label{tab:collections}
\end{table}

\myparagraph{Efficient Index Traversal}
To avoid the expensive scoring of all documents matching at least one query term in the document collection, several dynamic pruning techniques for disjunctive queries have been proposed. Safe dynamic pruning methods enrich the inverted index with additional information used to skip documents; during query processing, documents that cannot possibly score high enough to be in the list of the final top-$K$ documents are bypassed. 
Thus, the final list of returned documents is the same as in exhaustive scoring of all documents, but with significantly less work. 
The most widely adopted dynamic pruning techniques include \maxscore~\cite{maxscore}, \wand~\cite{wand}, \textsf{BlockMax} \wand\ (\bmw)~\cite{bmw}, and \textsf{Variable} \bmw~\cite{vbmw,mallia2019faster}.
In general, dynamic pruning techniques exhibit good efficiency speed-ups when able to exploit the underlying distribution of scores, global upper bounds, and local upper bounds~\cite{fntir2018}. Given a query, if few composing terms have high upper bounds, while the remaining posting lists or blocks of postings have low upper bounds, there are many possibilities to skip over documents that cannot make it into the final top-$K$. In particular, the {\bmtf} and {\sf{DPH}} {\cite{dph}} scoring functions are known to exhibit favorable score distributions, while scores based on probabilistic language models are less amenable to skipping {\cite{magicwand}}.

\begin{figure}[t]
\centering
\includegraphics[width=0.95\columnwidth]{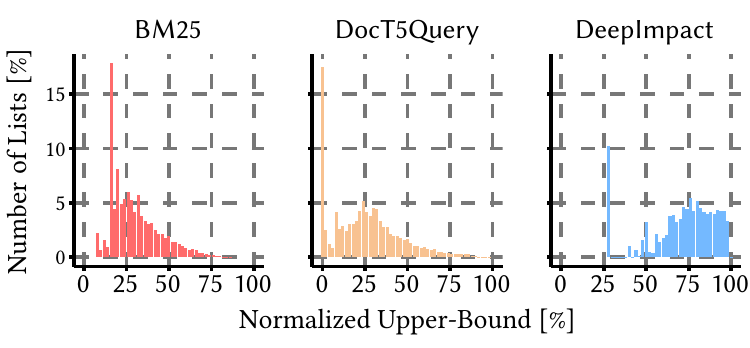}
\mycaption{
The distribution of list-wise upper-bound scores for all terms in the \msmarcodev. 
{\multibert} gives much higher weights to terms, on average, than the others.
\label{fig-dist}}
\end{figure}

\begin{figure}[t]
\centering
\includegraphics[width=0.85\columnwidth]{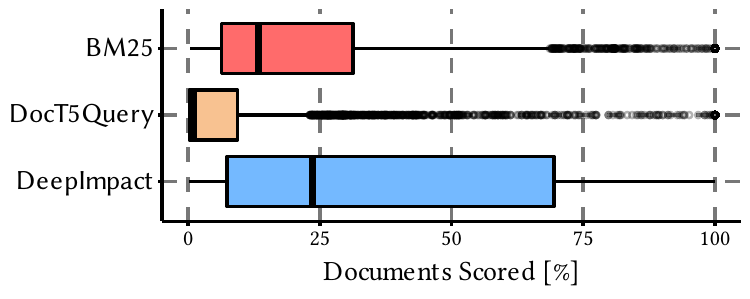}
\mycaption{
Relative percentage of documents scored (as compared to an exhaustive algorithm) with {\maxscore} for $k = 1000$
over the \msmarcodev.
\label{fig-postings}}
\end{figure}

\myparagraph{Motivation}
Prior work has empirically demonstrated that learned sparse models are typically slower than their traditional counterparts {\cite{deepimpactv1, deepct-efficiency, wacky}}, suggesting that the resulting score distributions are less amenable to dynamic pruning.
To illustrate this phenomenon, consider Figure~\ref{fig-dist}, which plots the list-wise upper-bound scores assigned to each of the query terms in the MSMARCO dev query set. 
Clearly, {\multibert} scoring yields many more high impact terms than {\bmtf} and {\doctfquery}, making it more difficult for dynamic pruning algorithms to avoid work. Figure~\ref{fig-postings} shows this effect, with {\multibert} scoring a much higher proportion of documents than {\bmtf} and {\doctfquery} under the popular {\maxscore} algorithm. 
What is the cause of these high impacts? Consider any common term such as {\emph{``the''}}; {\bmtf} will, by nature, give this term a low impact because it appears in most documents. However, the contextual nature of the language model {\deepimpact} is trained over means it can predict when {\emph{``the''}} is an important word, such as in an article discussing the definition or origin of the term {\emph{``the''}}. As such, we are interested in exploring heuristic processing techniques that can achieve the efficiency of the traditional rankers with the effectiveness of the learned sparse models.

\section{Fast Top-$K$ Guided Traversal}
\label{sec-double}

Motivated by the analysis in Section~\ref{sec:background}, we now propose a fast heuristic query processing strategy that exploits the favorable score distributions provided by {\bmtf}. 
In particular, our technique employs {\bmtf} scoring over a {\doctfquery} expanded index to decide which documents to process, but uses the {\multibert} model to compute the document scores, thereby running nearly as fast as plain {\bmtf} traversal, but with significantly better effectiveness. 
This method relies on the fact that {\multibert} uses {\doctfquery} to expand the underlying corpus and also uses word-level tokenization; thus the underlying posting lists are basically the same.\footnote{In practice, there may be slight differences due to different tokenization techniques.}
We hypothesize that the documents visited during index traversal over the {\doctfquery} index are typically high-impact documents under the {\multibert} model, which should allow fast, yet effective, retrieval.

\begin{figure}[t]
\centering
\includegraphics[width=0.95\linewidth]{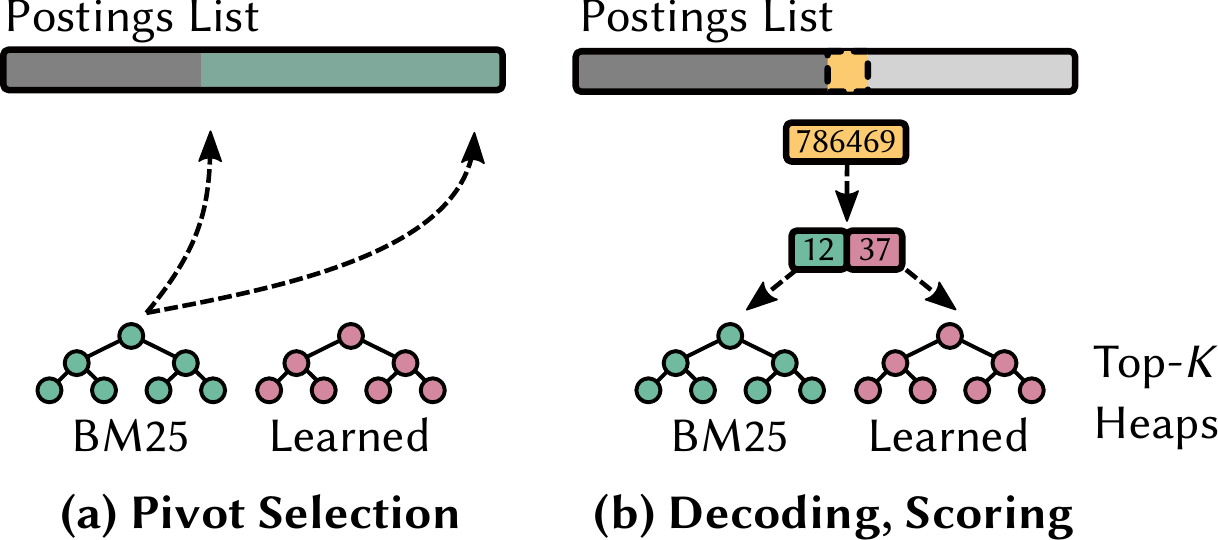}
\mycaption{Guided traversal for a single-term query: (a) the threshold of the {\bmtf} top-$K$ min-heap is used to find the next document to process, (b) the combined impact is decoded into its two component values, and checked against the respective results sets.
\label{fig-dual}
}
\end{figure}

To implement this query processing strategy, the impact value for both {\bmtf} (on the {\doctfquery} index) and {\multibert} must be stored for {\emph{each posting in the index}}.\footnote{Where the postings do not align one-to-one, the missing score is set to zero.}
Fortunately, due to the fact that a small number of bits is sufficient for representing the impact scores for these models, they can be packed into a single 32-bit integer, with each score represented with $16$ bits.
At query processing time, two top-$K$ min-heaps are initialized {\cite{next-page}}; the first heap maintains the top-$K$ documents according to {\bmtf}, and the second heap maintains the top-$K$ documents for {\multibert}. 
Query processing proceeds, as usual, using the {\bmtf} min-heap to drive the index traversal, including any dynamic pruning. 
When it comes time to score a document, the combined impacts are unpacked from the index, and each min-heap is updated.
At the end of processing, both top-$K$ heaps will contain the $K$ highest scoring documents that were visited during {\bmtf} traversal. 
Figure~\ref{fig-dual} sketches one cycle of this process, which we denote {\emph{guided traversal}} ({\sf{GT}}).

An interesting extension to the guided traversal process described above is to also account for {\emph{interpolated}} scoring
regimes {\cite{interpolate-dense}}. In particular, since the secondary learned top-$K$ set is computed over exactly those documents visited by the {\bmtf}-driven traversal, it is possible to compute interpolated scores in the secondary heap. We investigate this idea in our experiments by applying an unweighted linear interpolation between {\bmtf} and {\deepimpact}, denoted as {\emph{guided traversal with interpolation}} ({\sf{GTI}}).

While we expect that our {\gt} and {\gti} methods could generalize to other models, we only apply them to {\deepimpact} since \unicoil, \tildetwo, and \spladetwo\ use BERT WordPiece tokenization resulting in incompatible vocabularies (see Table~\ref{tab:collections}). We leave this investigation for future work.

\begin{table}[t]
\mycaption{Efficiency and Effectiveness for different processing strategies over the MSMARCO dev queries, with {\sf{GT}} and {\sf{GTI}} denoting our guided traversal without or with interpolation, respectively.}\label{tab:overall}
\centering
\begin{tabular}{l cccc}
\toprule
\multirow{2}{*}{Strategy} & \multicolumn{4}{c}{\msmarcodev}\\
\cmidrule(lr){2-5} 
& Mean & Median & $P_{99}$ & RR@10\\
\midrule
{\bmtf} & 5.7 & 4.3 & 24.0 & 0.187 \\
{\deepct} & 1.1 & 0.9 & 4.5 & 0.244\\
{\doctfquery}   & 3.8  & 2.8  & 16.5 & 0.272 \\
{\tildetwo} & 20.7 & 14.3 & 90.6 & 0.333\\
{\unicoil} & 37.9 & 21.4 & 194.5 & 0.352\\
{\spladetwo} & 219.9 & 201.3 & 581.6 & 0.369\\
{\multibert}    & 19.5 & 14.0 & 79.6 & 0.326 \\[1.1ex]
{\gt} & 4.8  & 4.2 & 16.1 & 0.326 \\
{\gti} & 5.0 & 4.4 & 16.7 & 0.341 \\
\bottomrule
\end{tabular}
\end{table}

\begin{figure}
\includegraphics[width=0.95\columnwidth]{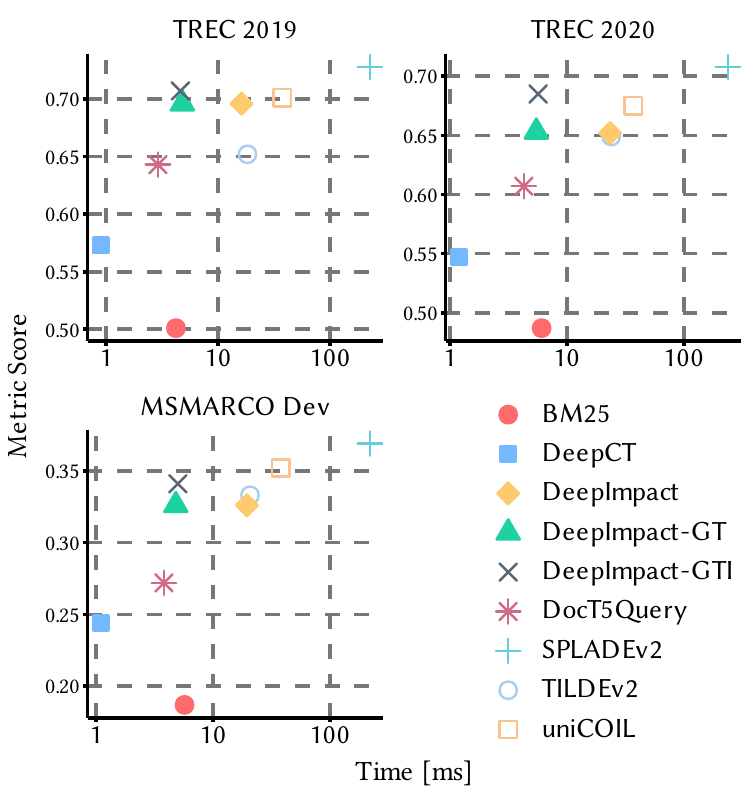}
\mycaption{Effectiveness vs Mean Latency overall collections. Both TREC 2019 and TREC 2020 are evaluated with NDCG@10, whereas the MSMARCO dev queries are evaluated with RR@10. Note the log scale on the $x$-axis, and the differing $y$-axis breaks.}
\label{fig-tradeoff}
\end{figure}

\section{Experiments and Results}\label{sec:exp}
In this section, we analyze the performance of the proposed method using the popular MSMARCO \cite{msmarco} dataset, consisting of 8.8M passages.
To evaluate query processing effectiveness and efficiency, all models are compared using the \msmarcodev{}, \trecdl{}~\cite{trec2019} and \trecdltw{}~\cite{trec2020} queries from the TREC Deep Learning passage ranking track. All experiments were conducted in memory using a single thread of a Linux machine with two 3.50 GHz Intel Xeon Gold 6144 CPUs and 512 GiB of RAM.

\myparagraph{Implementations}
We use Anserini~\cite{anserini} to generate the inverted indexes of the collections. We then export the Anserini indexes using the common index file format \cite{ciff}, and process them with PISA~\cite{pisa} using the {\sf{MaxScore}} query processing algorithm~\cite{maxscore}. All indexes are pre-quantized using 8-bit linear quantization to allow for ``sum of impact'' scoring {\cite{cclmt17wsdm}} and reordered using BP~\cite{dhulipala2016compressing,mackenzie2019compressing,mackenzie2021faster}.

\myparagraph{Ranking Models}
We use the \bmtf{} scoring method provided by Anserini with the recommended parameters $k_1 = 0.82$ and $b = 0.68$~\cite{pyserini}.
For \deepct{}, we  used the source code and data\footnote{\url{https://github.com/jmmackenzie/term-weighting-efficiency}} provided by \citet{deepct-efficiency}. \deepct{} also uses \bmtf{} to score documents, with $k_1 = 8.0$ and $b = 0.9$~\cite{deepct, deepct-efficiency}.
For \doctfquery{} we use the predicted queries available online,\footnote{\url{https://github.com/castorini/docTTTTTquery}} using 40 concatenated predictions for each passage in the corpus, as recommended by \citet{docTTTTTquery}; documents are scored with \bmtf{} using $k_1 = 0.82$ and $b = 0.68$.
For \unicoil\, we use the official implementation\footnote{\url{https://github.com/luyug/COIL}} which makes use of a {\doctfquery} expanded index. 
For \tildetwo\, we use the official implementation\footnote{\url{https://github.com/ielab/TILDE/tree/main/TILDEv2}} which first expands the whole MSMARCO passage collection with 200 expansion terms per document, and then indexes the expanded collection to generate an inverted index. This goes beyond the methodology adopted by \citet{tildetwo}, where {\sf{TILDE}}\ was used as a re-ranker. 
For \spladetwo\ we used the model provided by \citet{splade}.\footnote{\url{https://github.com/naver/splade}} We refer here to their best performing approach, also named {\sf{DistilSPLADE-max}}.
For \multibert{} we use the model provided by~\citet{deepimpactv1}.\footnote{\url{https://github.com/DI4IR/SIGIR2021}}

\myparagraph{Metrics}
To evaluate retrieval effectiveness, we follow common practice and report RR@10 for the MSMARCO dev queries and 
NDCG@10 for \trecdl{} and \trecdltw.
To carefully evaluate the retrieval performance of guided traversal with respect to the other strategies, we report the mean, the median, and the 99\,th percentile of the response time in ms. Values are taken as the mean of three independent runs.

\myparagraph{Results Analysis}
 In Table~\ref{tab:overall} we report our results on the large \msmarcodev, and in Figure~\ref{fig-tradeoff} we show the performance of the various strategies on all three query sets in terms of both effectiveness and efficiency, where the optimal region is the top left corner (minimal latency and maximal effectiveness).

Among the query processing strategies employing {\bmtf} as the underlying ranking model, namely \bmtf, \deepct, and \doctfquery, \deepct\ is the most efficient strategy, with a mean response time of $1.1$ms and a tail response time of $4.5$ms. However, the most effective strategy is \doctfquery\ across all query sets (see Figure~\ref{fig-tradeoff}). Among the baseline learned sparse models, namely \tildetwo, \unicoil, \spladetwo, and \multibert, \spladetwo\ is consistently the most effective one, but it is also the least efficient. With respect to the most efficient competitor, \multibert, its mean response time is $10$ to $14$ times larger, and its 99\,th percentile response time is $7$ to $8$ times larger, while \spladetwo\ effectiveness is 13.2\% higher on \msmarcodev, 4.6\% higher on \trecdl, and 8.4\% higher on \trecdltw.

Our proposed \gt\ strategy, deployed with {\multibert}, improves its mean response time by up to $4.3\times$, its median response time by up to $3.6\times$, and its tail response time by up to $5.8\times$, with no measurable impact on effectiveness. Our \gti\ strategy, interpolating {\bmtf} scores and learned {\multibert} impacts, slightly increases the mean response time of \gt\ by 0.2 ms, but it improves the overall effectiveness: +4.6\% on \msmarcodev, +1.6\% on \trecdl, and +4.9\% on \trecdltw. A pairwise $t$-test reported these improvements to be significant ($p \approx 10^{-10}$) for only the dev queries, with $p = 0.46$ and $p = 0.049$ reported for {\trecdl} and {\trecdltw}, respectively. 

Figure~\ref{fig-tradeoff} allows us to easily detect the characteristics of the several approaches and quickly identify the suitability of a method for a precise use case. We observe that \deepct{}, \doctfquery, \spladetwo, \gt{}, and \gti{} are on the Pareto frontier in all the plots, reducing the number of strategies to choose from depending on the efficiency-effectiveness constraints.

\section{Conclusions}
Learned sparse models result in substantial retrieval quality improvements while reducing the efficiency gap between neural retrieval and the faster traditional sparse models based on inverted indexes. This gap correlates with vocabulary structures, document expansion techniques, and query expansion strategies, making the several learned sparse models quite different efficiency-wise. %
In this work, we have proposed a ``guided traversal'' approach to accelerate index traversal by coupling a learned sparse ranking model with a traditional ranking model. Our proposed approach employs {\bmtf} ranking over a {\doctfquery} expanded index to {\emph{lead}} the index traversal, but uses the {\multibert} ranking impacts to compute document scores.
Our preliminary results on top of \multibert\ show that our guided traversal approach is almost able to match the processing efficiency of traditional sparse models, while also improving the retrieval effectiveness of the learned sparse models through interpolation of the scores of the traditional and learned sparse ranking models. In future work, we plan to explore whether our guided traversal heuristic is practical for other learned sparse models, as well as different efficient query processing strategies.

\myparagraph{Software}
In the interest of reproducibility, software is available for generating our guided traversal runs.
See {\url{https://github.com/DI4IR/dual-score}} for more information. 

\myparagraph{Acknowledgements}
This research was partially supported by the Australian Research Council Discovery Project DP200103136. We thank Jimmy Lin and Xueguang Ma for their assistance with the experimental resources and Matthias Petri for his helpful discussions related to this work.

\balance
\bibliographystyle{abbrvnat}
\bibliography{www2022}

\end{document}